# Remote Electric Powering by Germanium Photovoltaic Conversion of an Erbium-Fiber Laser Beam

**Richard Soref** [1], **Francesco De Leonardis**[2], **Oussama Moutanabbir**[3], **and Gerard Daligou**[3]

[1] Engineering Department, University of Massachusetts at Boston, Boston, MA 02125 USA

[2] Department of Electrical and Information Engineering, Politecnico di Bari, Bari, Italy

[3] Department of Engineering Physics, Ecole Polytechnique de Montreal, Montreal, Quebec, Canada

E-mails: richard.soref@umb.edu (R. Soref), Francesco.deleonardis@poliba.it (F. De Leonardis) oussama.moutanabbir@polymtl.ca (O. Moutanabbir), gerard.daligou@polymtl.ca (G. Daligou)

**ABSTRACT**

The commercially available 4000-Watt CW Erbium-doped-fiber laser, emitting at the 1567-nm wavelength where the atmosphere has high transmission, provides an opportunity for harvesting electric power at remote "off the grid" locations using a multi-module photovoltaic (PV) "receiver" panel. This paper proposes a 32-element monocrystalline thick-layer Germanium PV panel for efficient harvesting of a collimated 1.13-m-diam beam. The 0.78-m$^2$ PV panel is constructed from commercial Ge wafers. For incident CW laser-beam power in the 4000 to 10000 W range, our thermal and electrical and infrared simulations predict 660 to 1510 Watts of electrical output at panel temperatures of 350 to 423 Kelvin.

## INTRODUCTION

In the emerging field of "power by light" [1,2], also known as laser power transmission (LPT) [3] and optical wireless power transfer [4-15], the experiments reported in the literature utilized a continuous-wave (CW) laser-beam output power of 50 Watts, or less. This prior art does not include recent advances in the CW short-wave infrared (SWIR) laser art. The present paper targets lasers that emit in the SWIR defined here as the 900 to 1700-nm wavelength region. In addition, the present paper considers only the commercially available lasers whose CW output power is 4000 Watts or more, lasers that we define as ultra-high power (UHP) laser transmitters. A scan of the laser-development literature and a survey of reports from laser manufacturers reveals that there are currently only two UHP SWIR lasers. They are: the Ytterbium-doped fiber laser (YFL) emitting at the 1075-nm center wavelength, and the Erbium-doped fiber laser (EFL) emitting at the 1567-nm wavelength. The beams of both lasers travel with very low loss through the earth's atmosphere because of the atmosphere's transparency for those waves.

In this UHP context, a recent theoretical-physics paper by the present authors [16] predicted efficient optical-to-electrical (OE) conversion, or "harvesting," of the YFL beam by a multi-module silicon photovoltaic (PV) "panel," and [16] is, to the best of our knowledge, the first UHP-beam-harvesting paper. The present paper is the second UHP-OE-conversion paper, and this present work is motivated by the fact that the silicon PV panel of [16] cannot and will not harvest the EFL beam because the 1.12 eV silicon bandgap energy is well above the 0.79 eV EFL photon energy. An alternative PV is needed, and an elemental semiconductor PV is the simplest and perhaps-best choice. In view of that, a new aspect of this paper is that we propose an optimum semiconductor PV panel to harvest the EFL; specifically, we propose-and-analyze the mono-crystalline thick-layer Germanium PV panel to provide efficient-and-effective EFL OE conversion. This is a new Ge panel, not discussed in the literature. The literature instead reports thin-film and heterojunction Ge modules [17-24]. We show simulations of panels having modules fabricated from 300-mm commercial Ge wafers. Our numerical modeling includes large-area series-connected panels that harvest a collimated large-area EFL beam. An additional new aspect here is that we quantify the Ge PV panel's harvesting of the YFL, which is an added benefit of the Ge approach. The EFL mission is primary here, while the YFL harvesting is secondary but quite useful.

The sections of this paper include a background discussion, design of the encapsulated Ge PV-diode module, thermal-and-electrooptical simulations of this module, numerical results for the conversion efficiency of Ge+EFL and Ge+YFL, a comparison to Si +YFL conversion, the optimized 32-module Ge panel, Ge TPV, and the Ge+EFL system offering both RF communications and DC electric output.

## BACKGROUND DISCUSSION

Since the EFL photon energy is above the 0.66 eV Ge bandgap, the Ge+EFL system offers efficient conversion and large electric output. Because several commercial vendors offer oriented crystal Ge wafers in diameter up to 300 mm, this paper proposes the use of large-area 156 mm x 156 mm monocrystal Ge PV modules that are fabricated from those wafers. Those modules are then grouped into a planar Ge PV "panel," such as a 32-element array that gives good overlap with the large-diameter of the collimated, incident laser beam. It is an array in which the constituent modules are interconnected electrically in series. This panel is quite analogous to a Si solar panel in terms of its layout and its construction or "encapsulation." Another Ge-to-Si PV similarity is the utilization of "thick layers" in each Ge PN junction, specifically a Ge PN diode whose total thickness is 180 μm is used here. A motivation here is to reach kilowatt electric output using commercial EFLs. There is an EFL available at the 4000 W CW level [25], and this output could in principle be increased to 10000 W CW in a custom product.

Regarding system details, in order to obtain targeting, the EFL would be mounted in an aiming device (for example, mounted in a tilt/rotation turret on top of a truck) so that the beam can be directed to impinge upon the remotely located PV receiver. The laser can be ship-borne or space-borne or airborne or on-ground as described.

There are numerous remote-powering applications such as the eleven scenarios listed in [16] (all of which apply here) where the PV-panel receiver is in an "off the grid" location, and where fuel-powered portable electric generators are not readily available or are not desired. If the PV electricity is not immediately used, the laser power transmission (LPT) can be used to charge storage batteries.

We should state at the outset that the Ge laser power system (and such systems generally) face barriers to their widespread adoption. These challenges are in the areas of economics and safety. In particular, the system cost is high due to capital cost of the UHP laser, and because Ge PV is more costly than the Si PV. When considering the energy that must be supplied to actuate the laser, the overall energy efficiency (i.e., energy-in/energy-out) of the LPT might be low. There are issues of eye safety at transmitter and receiver, and for the case of 20 kW lasers, the high operation temperature of the PV panel might be hazardous. However, there are situations/applications where the LPT benefits outweigh the disadvantages.

## THERMAL and ELECTRICAL-OPTICAL SIMULATIONS

**Photovoltaic Module Design** For the optical wireless system in [16], we adopted the multi-material "encapsulation" architecture of the commercial monocrystal silicon solar cell for the Si PV module. Because of the success of that approach, we have decided to utilize here the same multi-material construction for the LPT Ge PV module, as is illustrated in Figure 1. In Figure 1, if we define *t* as the "optimum" overall thickness of the mostly-P- region plus N-region, there is a design rule for determining $t$, which is $t = 1/\alpha$, where $\alpha$ is the room-temperature absorption coefficient of the dominant P-type Ge at the 1567-nm wavelength. Turning to the Ioffe reference compilation [26] we find that $\alpha$ = 55 cm$^{-1}$, yielding $t$ = 180 μm. This is a layer thickness that is generally much higher than that produced by Ge epitaxial growth procedures, which implies that the Fig.-1 diode is simply fabricated from a P-type Ge commercial wafer.

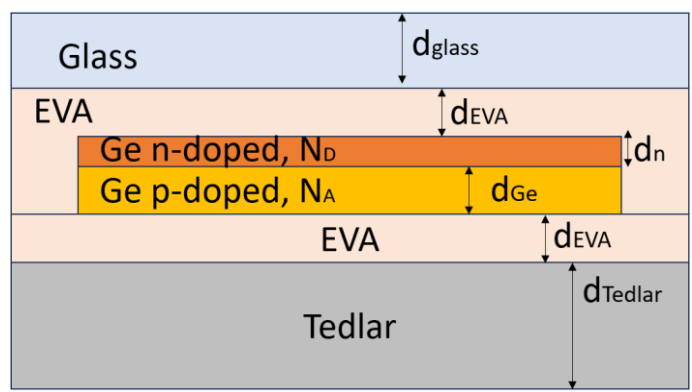

**Fig. 1** | Cross section view of Ge PV cell utilizing the commercial Si solar cell architecture. Relevant parameters and materials are indicated. The cell has an area of 15.6 cm x 15.6 cm. For the n-doped layer: dn=0.2 μm, $N_D$=10$^{19}$ cm$^{-3}$. For p-doped layer: $d_{Ge}$=180 μm, $N_A$=10$^{17}$ cm$^{-3}$.

**Numerical Results** A general theoretical approach to simulating the thermal and optical-electrical responses of the large-area PV module was presented in [16] including six equations that quantified the PV performances. We have adopted here for Ge the same modeling-and-simulating framework, the same equations, and have taken from the literature [26] the necessary Ge experimental parameters, such as the infrared absorption coefficients versus wavelength and bandgap versus temperature [26]. The infrared absorption coefficient and the Ge bandgap narrowing effect [26] have been fitted by means of Eq. 18 of Ref. [27] and Eq. 1 of Ref. [28], respectively. Moreover, carrier mobilities and effective masses have been evaluated using the diffusion coefficients and the effective density states given in Table I and II of [29]. Using this framework, we made a set of calculations whose results are given in Figures 2 – 5 as follows.

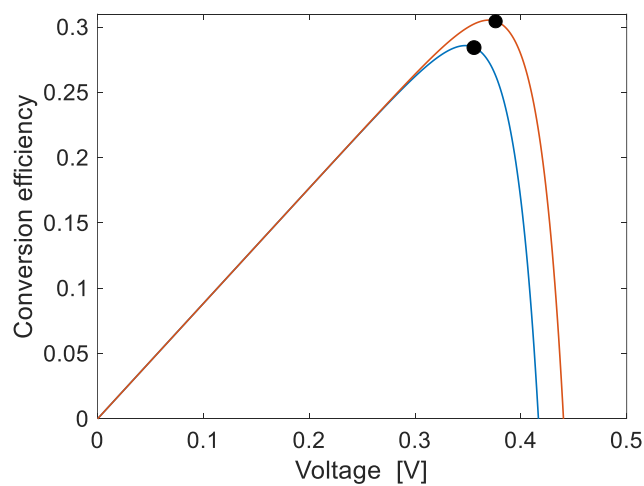

**Fig. 2** | Calculated OE conversion efficiency as a function of the Ge PV voltage, for input Er laser power values of 4,000 W and 10,000 W and $d_{Ge}$ =180 μm, respectively. In the simulations, the laser emission wavelength and the laser spot radius

($r_{sp}$) are 1567 nm and 80 cm, respectively. The operative temperature is forced at 300 K. These are the parameters used for simulating the Fig.-1 thermal responses.

**Table 1 | Physical Parameters of Thermal Simulations.**

| Parameters | Materials | | | |
|---|---|---|---|---|
| | **Germanium** | **Glass** | **EVA** | **Tedlar** |
| Density [kg/m$^3$] | 5323 | 2450 | 950 | 1200 |
| Thermal conductivity [W/mK] | 60 | 2 | 0.311 | 0.15 |
| Heat capacity at constant pressure [J/kgK] | 310 | 500 | 2090 | 1250 |

It is assumed here that there is a finned metal heat sink contacting the rear face of the PV in order to provide passive PV cooling, which means cooling via convection of ambient air.

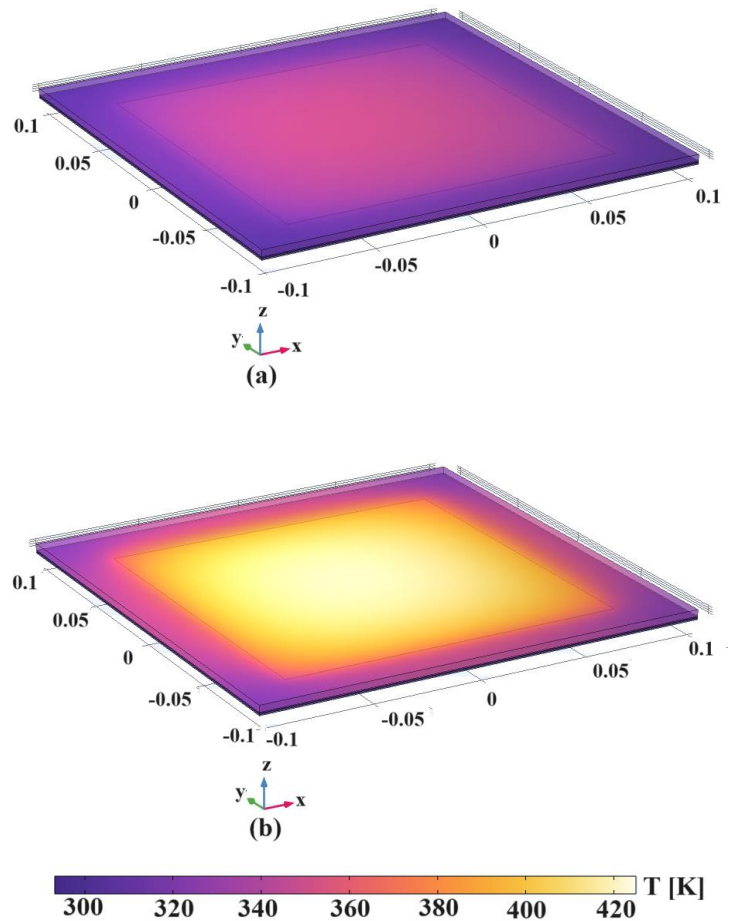

**Fig. 3 |** Calculated temperature distribution in a 156 mm x 156 mm germanium PV module. (a) Laser power 4000 W and $d_{Ge}$ =180 μm; (b) Laser power 10000 W and $d_{Ge}$ =180 μm. In the simulations, the laser emission wavelength, the emission bandwidth and the laser spot radius ($r_{sp}$) are 1567 nm, 10 nm and 80 cm, respectively. The local yellow or orange or red or

purple color shown here in the two drawings signifies the local temperature of the Ge NP "plate" as is specified by the Kelvin-versus-color scale given here.

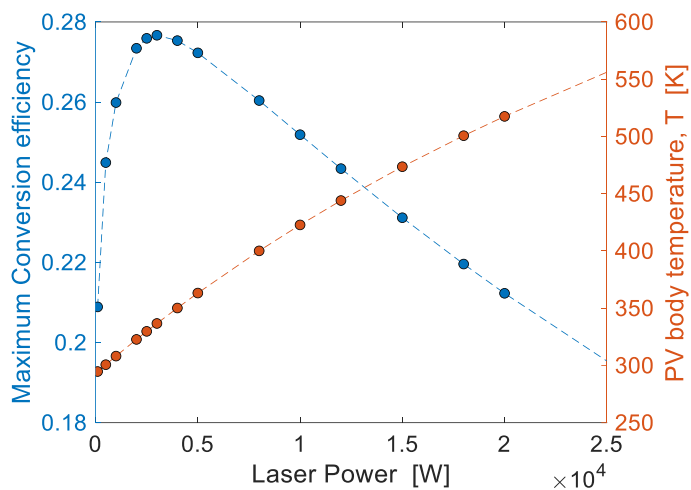

**Fig. 4 |** Maximum conversion efficiency and Ge PV body temperature as a function of the incoming CW Er laser power ranging from 100 W to 25000 W and for $d_{Ge}$ =180 μm. In the simulations, the laser emission wavelength, the emission bandwidth and the laser spot radius ($r_{sp}$) are 1567 nm, 10 nm and 80 cm, respectively.

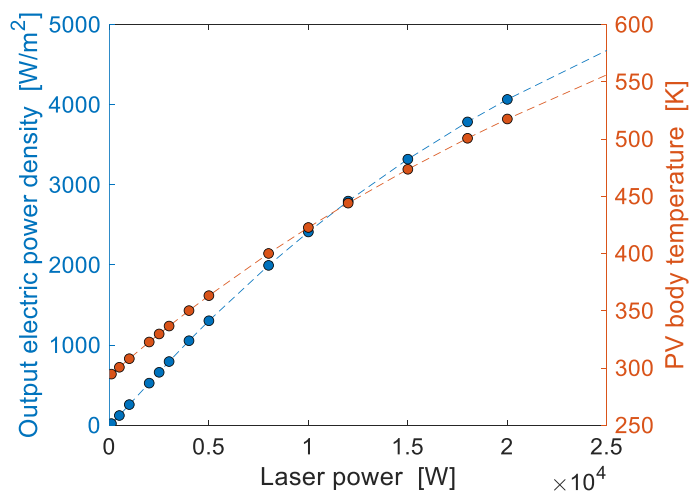

**Fig. 5 |** Output electrical power density of Ge PV cell and its associated PV body temperature as a function of the EFL CW input-to-PV power, ranging from 100 W to 25000 W and for $d_{Ge}$ =180 μm. In the simulations, the laser emission wavelength, the emission bandwidth and the laser spot radius ($r_{sp}$) are 1567 nm, 10 nm and 80 cm, respectively.

Regarding the OE conversion efficiency calculated in Figures 2 and 4, the Gaussian shape of the beam laser has been taken into account by applying Eq. 4 of Ref. [16]; that is, for a laser beam having a Gaussian spatial distribution, the laser flux on the top PV surface is given by:

$$Flux = (1-R)\frac{2 \cdot P_{laser}}{\pi \cdot r_{sp}^2} e^{\left(-2\frac{(x-x_{focus})^2 + (y-y_{focus})^2}{r_{sp}^2}\right)}$$

where $R$ is the reflectivity at the top glass surface, $P_{laser}$ is the laser power arriving at this PV surface and $r_{sp}$ represents the laser spot radius (where the beam intensity has fallen down to $1/e^2$ of the central beam power). The terms $x$ and $y$ are the coordinates in the xy plane of the PV cell and $x_{focus}$ and $y_{focus}$ represent the spot's center point. In turn, the laser Flux is used to characterize the heat source in each PV layer, according to the Beer-Lambert law, given by Eq. (2) of Ref. [16]. In our simulations we have assumed $r_{sp}$=80 cm, and we have considered the effective area of the beam $A_b = \pi r_e^2$, where $r_e$ is the effective beam-spot radius given by $r_e = r_{sp}/\sqrt{2}$, yielding $A_b = \pi r_{sp}^2/2$. Moreover, our investigations indicate we that the maximum OE conversion efficiency weakly decreases with increasing spot radius.

In Figure 4, we observe an initial rise in maximum conversion efficiency as the laser power is increased, and subsequently a decline-in-efficiency with further laser increase. We can account for this behavior by taking into account "thermal aspects" as follows: Under monochromatic illumination, the temperature dependence of the conversion efficiency induces the existence of $T_{peak}$ to which corresponds a peak in the maximum conversion efficiency. The existence of this peak is induced by the complex temperature dependence given by Eqs. (2), (3), and (6) of Ref. [16]. In the present system, as the PV temperature rises up from 293K towards 550K, two opposite trends are recorded. On the one hand, the increase in laser power induces an efficiency decrease, while on the other hand the increase produces an increase in the temperature that in turn increases the Ge absorption coefficient (and then the PV generated density current) because of the Ge bandgap narrowing. Moreover, our investigations indicate that $T_{peak}$ decreases by decreasing the laser emission wavelength ($T_{peak}$ is lower than 293 K for $\lambda <$ 920 nm). In this context, the laser heating process will induce a reduction of the maximum conversion efficiency as the PV body temperature increases above 293 K, similar to what happens to the PV during sun illumination [30].

Figure 5 is the principal result of this paper, showing high 1055 to 2412 W/m² electric output densities over the EFL incident-beam 4000 to 10000 W range mentioned above, operating with 350K to 423K PV body temperatures.

The laser power that is cited in Figures 4 and 5 is the power that impinges upon the top (front) surface of the PV module. If the beam from the laser transmitter travels a "long distance" through the earth's atmosphere to reach the PV receiver, the effect of the atmosphere upon power transfer [31] can be quantified by turning to Figure 1 of [16] where the transmittance of the 1.567 μm EFL is shown as 77% radiative transfer across one nautical mile-of-travel at sea level. If we choose, for example, to locate the beam harvester two miles away from the laser source, then the power reaching the PV receiver becomes 59% of the source output.

We discussed earlier the additional YFL conversion offered by Ge PV, and here we have quantified that performance in order to supplement the EFL results.   Our numerical modeling of this Ge+YFL OE is presented here in Table 2 along with Ge+EFL. Because the YFL beam conversion was investigated in [16] where Si PV was the receiver, we decided to add in this same Table 2 several Si+YFL results to provide a comparison with the Ge+YFL predictions. To make this a fair comparison, we impose the constraint in Table 2 that the Si PV cells must have the same 180 μm thickness as the Ge PV cells. However, the Table does not reveal the full or optimum Si+YFL performances because much higher output power densities were found [16] when increasing the Si cell thickness $t$ from 180 μm to 500 μm.

**Table 2 | Summary of Results.**

| PV and FL | Metrics | | | | |
|---|---|---|---|---|---|
| | CW laser power output [W] | PV body temperature [K] | Efficiency [fractional] | Output electric power density [W/m$^2$] | Output electric power [W] For A=0.0243 m$^2$ |
| Ge+EFL 1567 nm | 4000 | 350 | 0.27 | 1055 | 26 |
| Ge+EFL 1567 nm | 10000 | 423 | 0.25 | 2412 | 59 |
| Ge+EFL 1567 nm | 20000 | 518 | 0.21 | 4056 | 99 |
| Ge+YFL 1075 nm | 4000 | 354 | 0.18 | 680 | 17 |
| Ge+YFL 1075 nm | 10000 | 427 | 0.16 | 1533 | 37 |
| Ge+YFL 1075 nm | 20000 | 520 | 0.13 | 2550 | 62 |
| Si+YFL 1075 nm | 4000 | 320 | 0.13 | 530 | 13 |
| Si+YFL 1075 nm | 10000 | 401 | 0.26 | 2546 | 62 |
| Si+YFL 1075 nm | 20000 | 564 | 0.31 | 6272 | 152 |

    YFL's are already available at extremely high CW output power. On the idea that the EFL might someday be extended to 20000 W, we also decided to determine the 20000 W CW EFL responses of Ge PV along with the 20000 W CW Ge+YFL behavior. All of this is summarized in this presentation of efficiency, cell temperature, output power density and output electric power.

    An examination of this Table shows that Ge offers unique and practical, high-performance EFL conversion. As to YFL conversion, the Ge performances are "mixed," which means that the Ge+YFL efficiencies are lower than those for Ge+EFL, but are still useful and valuable.   However, it is also clear that silicon generally "dominates" in the YFL-conversion category.

    If we ask about factors that contribute to the predicted lower Ge+YFL conversion, we note that the reflectivity of the top glass surface in Figure 1 is lower at the 1567-nm EFL wavelength than at the 1075-nm YFL wavelength. This means that the Ge+YFL heat source is somewhat higher in the above Flux equation, and that reduces the efficiency.

    The Table 2-results pertain to an individual PV module for which we took into account the external radiative efficiency (ERE), a parameter that accounts for carrier recombination losses. Resistive losses were not included. However, resistive effects in individual "solar type" PV cells do reduce the fill factor FF and then reduce the efficiency of the cell by dissipating power in the resistances. In particular, the most common parasitic resistances are the series resistance and the shunt resistance. In our PV modules, typical values for area-normalized series and shunt resistances are around 0.5 Ωcm$^2$ and in the MΩcm$^2$ range, respectively. In this context, due to the large area used for one cell (0.0243 m$^2$), the dominant resistive effect is determined by the series resistance R$_{SER}$. To accurately represent our case, we shall make the assumptions $d_{Ge}$=180 μm, laser power of 4000 W and PV body temperature of 350 K. As evidenced in [32], the series and shunt resistances for a Ge PV cell decrease as the incident power density increases. In the thermophotovoltaic approach proposed in [32], the authors measure the series and shunt resistance, respectively, of 0.02 Ωcm$^2$ and 0.22 kΩcm$^2$ during spectral irradiance at 2000 K blackbody temperature. Since in

our case the operative input power ranges from 4000 W (0.40 W/cm$^2$) to 10000 W (0.99 W/cm$^2$), and according to Table 2 of [32] it is valid to assume that the series and shunt resistance change from 0.31 Ωcm$^2$ to 0.11 Ωcm$^2$ and from 1.55 kΩcm$^2$ to 0.80 kΩcm$^2$, respectively. For that reason, the electric power loss is estimated around 37% for the single PV cell.

## MULTI-MODULE Ge PV PANEL

If we now construct a planar panel from an array of cells, and if we assume that the panel is illuminated by the same cylindrical collimated EFL beam having 1.00 m$^2$ effective area and 0.564 m effective spot radius that is used for the individual PV module of 0.0243 m$^2$ area, then the area of the panel enters into the estimate of its electrical output power, and that output scales in proportion to panel area.   We now propose that the 32-module array illustrated in Figure 6 is an optimum choice because of its beam-filling geometry and its 0.78 m$^2$ area.   One approach to determine the panel's output is to simply multiply the Table-2 module outputs by the area-increase-factor of 32; but to be conservative, we shall instead use 80% of the "area factor" because the infrared intensity is distributed nonuniformly across the 1 m$^2$ beam "spot."   Thus 80% of 32 is 25.6 and that is the factor used here to find the panel outputs, which are 666 W (at 350K panel) and 1510 W (at 423K panel) for 4000 W and 10000 W beams, respectively. An important aspect of the panel is the electrical interconnection of cells that is chosen. As we discussed in [16], it is essential to connect the 32 modules in series, and when that is done, the series-resistance of each cell induces only a negligible electric power loss in the panel.

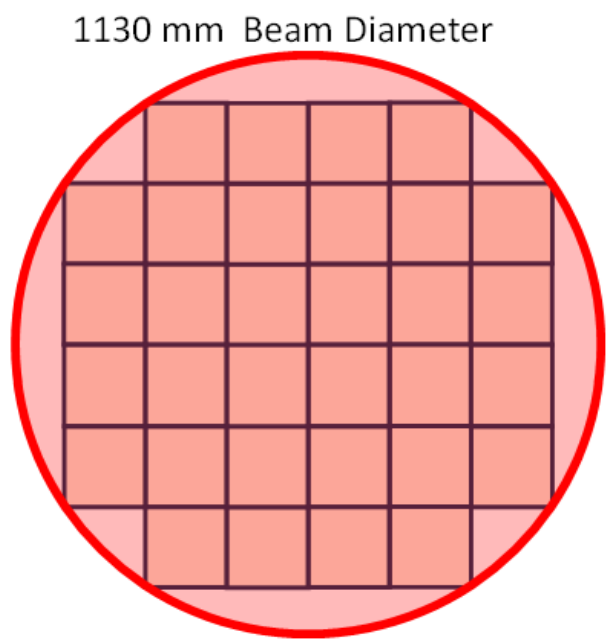

**Fig. 6 |** Proposed Ge PV panel comprised of 32 identical interconnected PV modules in a layout offering strong overlap with the 1567-nm collimated beam.

We should note in passing that Figure 6 has immediate application to thermo-photovoltaic TPV sensing. For example, [32] has quantified OE conversion of 9.7 to 11.8% OE for the 2000K filtered blackbody radiation intensity of 3.66 to 36.6 W/cm$^2$ incident upon the Ge TPV NP diode. For our 0.78 m$^2$ Ge TPV panel, this TPV result translates immediately to a DC electric output of 2769 W to 33687 W from the Fig.-6 panel.

We want to add that there are potential alternative semiconductor materials (alternatives to Germanium) that could be utilized to construct the EFL-harvesting PV panel. However, those PV- material choices have some disadvantages. To be specific, the compound semiconductor GaSb has $E_g$ = 0.68 eV at 300K, while InN has $E_g$ = 0.65 eV at 300K, and both of these PV materials could efficiently convert the 1567-nm EFL, but there are issues of GaSb/InN wafer-size availability and wafer cost. In principle, the PV panel could be constructed from InGaAs or InGaAsP alloy whose alloy composition would be chosen to give absorption α at 300K of 50 to 100 cm$^{-1}$ at the EFL wavelength, according to the $t \sim 1/\alpha$ design rule. In that case, the 100-to-200-μm PV layers required would present very severe epitaxy challenges for the alloys. Regarding the narrow-bandgap semiconductors InAs, InSb, PbSe, and PbS (all of whose room-temperature α is "high" at 1567 nm), these materials are probably not viable to use for the PV panel because the 1/α approach gives a sub-micron PV NP layer, which implies that the panel will heat up to an unmanageably high temperature during UHP illumination.

Returning to the Ge panel, we should mention that there is the possibility of attaining high-speed optical communications [33,34] simultaneously in conjunction with the DC energy harvesting; a dual DC+RF function of the system, operating along the lines proposed in [35]. In the present case, a small-area radio-frequency (RF) electrooptical modulator (external to the CW laser) would be placed within the laser's two-lens collimating optics to intercept the beam in a focal region where the beam diameter is relatively small. That modulator [36,37] would put RF information on the beam with a small depth-of-modulation, and then at the remote receiver, a small-area RF photodetector would be placed at the rear face of the large PV panel to "sample" the small amount of the infrared beam that is transmitted through the PV panel (the "residual transmission" of the panel), and that photodetector [38,39] would demodulate the intensity-variations, giving the signals that were put onto the laser beam. Both RF devices could use Ge or SiGe as their active material [40].

## CONCLUSIONS

We have proposed and analyzed the efficient, directed-beam, electric-energy harvesting system comprised of the 1567-nm UHP CW Er-doped fiber laser source and the monocrystal Germanium 32-PN-diode-module series-connected photovoltaic "panel" receiver that is sited a long distance away from the collimated and directed laser. The theoretical modeling results are favorable because 666 W to 1512 W of electric power generation is predicted for 4 kW to 10 kW incident beaming, with the passively heat-sinked panel rising to 350 to 423 K during operation. An added benefit is that the same Ge panel gives 435 W to 947 W electric harvesting of the 1075-nm Yb-doped 4 kW to 10 kW fiber lasers.

## REFERENCE

1. Algora, C; Garcia, I; Delgado, M; Pena, R; Vazquez, C; Hinojosa, M; Rey-Stolle, I, Beaming power: Photovoltaic laser power converters for power-by-light, Joule, 2022, 6, 340-368.
2. Gou, Y; Mou, Z; Wang, H; Chen, J; Wang, H; Deng, G, High-performance laser power converters with resistance to thermal annealing, Optics Express, 2024, 32(5), 8335.
3. Liu, H; Zhang, Y; Hu, Y; Tse, Z; Wu, J, Laser-power transmission and its application in laser-powered electrical motor drive, Power Electronics and Drives, 2021, 6(41), 167.
4. Haas, H; Elmirghani, J; White, I, Optical wireless communication, Philosophical Transactions Royal Society, 2020, A378, 2020051.
5. Javed, N; Nguyen, N; Naqvi, S; Ha, J, Long-range wireless optical power transfer system using an EDFA, Optics Express, 2022, 30(19).
6. Zheng, Y; Zhang, G; Huan, Z; Zhang, Y; Yuan, G; Li, Q; Ding, G; Lv, Z; Ni, W; Shao, Y; Liu, X, Wireless laser power transmission: Recent progress and future challenges, Space Solar Power and Wireless Transmission, in press 2024.
7. Balaji, C, Wireless laser power transmission: A review of recent progress, International Journal of Engineering Research

& Technology, 2018, 6(14).

8. W. Xu, X. Wang, W. Li and C. Lu, Research on test and evaluation of laser wireless power transmission system, EURASIP Journal on Advances in Signal Processing, 2022, 2022:20.

9. Mohammadnia, A; Ziapour, B; Ghaebi, H; Khooban, H, Feasibility assessment of next-generation drones powering by laser-based wireless power transfer, Optics and Laser Technology, 2021,143, 107283.

10. He, T; Zheng, G; Wu, Q; Huang, H; Wan, L; Xu, K; Shi, T; Lv, Z, Analysis and experiment of laser energy distribution of laser wireless power transmission based on a powersphere receiver, Photonics MDPI, 2023, 10, 844.

11. Zhou, W; Jin, K, Power control method for improving efficiency of laser-based wireless power transmission system," IET Power Electronics, 2020, 13(10), 2096.

12. Kim, G; Park, Y, Maximizing wireless power transfer for electric vehicles with high-intensity laser power beaming and optical orthogonal frequency division multiplexing, Transportation Research Proceedings, 2023,70.

13. Yao, D; Gao, B; Qiang, H; Wang, X; Wen, K; Wang, D, Laser wireless power transfer and thermal regulation method driven by transient laser grating, AIP Advances, 2022, 12, 105001.

14. Jin, K; Zhou, W, Wireless laser power transmission: Review of recent progress, IEEE Transactions on Power Electronics, 2019, 34, (4) 3842.

15. Zhang, Q; Fang, W; Liu, Q; Wu, J; Xia, P; Yang, L, Distributed laser charging: A wireless power transfer approach, arXiv:1801.03835v3 [eess.SP] 9 Oct 2018.

16. Soref, R; De Leonardis, F; Daligou, G; Montanabbir, O, Directed high-energy infrared laser beams for photovoltaic generation of electric power at remote locations, APL Energy, 2024, 2(2).

17. Osterthun N; Neugebohrn N; Gehrke K; Vehse M; Agert C, Spectral engineering of ultrathin germanium solar cells for combined photovoltaic and photosynthesis, Optics Express, 2021, 29(2), 938-950.

18. Lombardero I; Ochoa M; Miyashita N; Okada Y; Algora C, Theoretical and experimental assessment of thinned germanium substrates for III–V multijunction solar cells, Progress in Photovoltaics: Research and Applications, 2020, 28(11) 1095-1214.

19. Hekmatshoar B; Shahrjerdi D; Hopstaken M; Fogel K; Sadana D, High-efficiency heterojunction solar cells on crystalline germanium substrates, Applied Physics Letters, 2012, 101, 032102.

20. Fernandez J; Janz S; Suwito D; Oliva E; Dimroth F, Advanced Concepts for High-Efficiency Germanium Photovoltaic Cells, 33rd IEEE Photovoltaic Specialists Conference. 2008.

21. Sun G; F Chang F; Soref R, High efficiency thin-film crystalline Si/Ge tandem solar cell, Optics Express, 2010, 18(4), 3746-3753.

22. Azizman M; Azhari A; Ibrahim N; Halin D; Sepeai S; Ludin N; Nuzaihan N; Nor M; Ho L, Mixed cations tin-germanium perovskite: A promising approach for enhanced solar cell applications, Heliyon, Cell Press, 2024, 10(8), e29676.

23. Zhu X; Cui M; Wang Y; Yu T; Li Q; Deng J; Gao H, Evaluation of electricity generation on GeSn single-junction solar cell, International Journal of Energy Research, 2022, 46(10).

24. Zhou Z; Liu W; Guo Y; Huang H; Ding X, Design Simulation and Optimization of Germanium-Based Solar Cells with Micro-Nano Cross-Cone Absorption Structure, Coatings, 2022, 12(11), 1653.

25. IPG Photonics CW laser series ELS-4000.

26. Germanium physical parameters listed at the website: https://www.ioffe.ru

27. Tran, H; Du, W; Ghetmiri, S; Mosleh, A; Sun, G; Soref, R; Margetis, J; Tolle, J; Li, B; Naseem, H; Yu, S, Systematic study of $Ge_{1-x}Sn_x$ absorption coefficient and refractive index for the device applications of Si-based optoelectronics, Journal of Applied Physics, 2016, 119, 103106.

28. Chang, G; Yu, S; Sun, G, "GeSn Rule-23"-The Performance Limit of GeSn Infrared Photodiodes, Sensors MDPI, 2023,

23, 7386.

29. Baran, V; Cat, Y; Sertel, T; Ataser, T; Sonmez, N; Cakmak, M; Ozcelik, S, A Comprehensive Study on a Stand-Alone Germanium (Ge) Solar Cell," Journal of Electronic Materials, 2020, 49(2).

30. Singh P; Ravinda N, Temperature dependence of solar cell performance—an analysis, Solar Energy Materials & Solar Cells, 2012, 101, 36–45.

31. Fahey T; Islam M; Gardi A; Sabatini R, Laser Beam Atmospheric Propagation Modelling for Aerospace LIDAR Applications, Atmosphere MDPI, 2021, 12, 918.

32. Gamel, M; Ker, P; Rashid, W; Lee, H; Hannan, M; Jamaludin, M, Performance of Ge and $In_{0.53}Ga_{0.47}As$ Thermophotovoltaic Cells under Different Spectral Irradiances, IEEE Access, 2021, 9, 37091-37102.

33. Garg D; Nain A, Next-generation optical wireless communication; a comprehensive review, Journal of Optical Communications, De Gruyter, 2021.

34. Walsh S; Karpathakis S; McCann A; Dix-Matthews B; Frost A; Gozzared D; Gravestock C; Schediwy S. Demonstration of 100 Gbps coherent free-space optical communications at LEO tracking rates, Scientific Reports, 2022, 12, article number: 18345.

35. Kessler-Lewis, E; Polly, S; Hubbard, S; Hoheisel, R, Demonstration of a monolithically integrated hybrid device for simultaneous power generation and data modulation, IEEE Journal of Photovoltaics, 2024, 14 (2).

36. Soref R; Sun G; Cheng H, Franz-Keldysh electro-absorption modulation in germanium-tin alloys, Journal of Applied Physics, 2012, 111 (12) 31.

37. Bennett B; Soref R, Analysis Of Franz-Keldysh Electro-Optic Modulation In InP, GaAs, GaSb, InAs, And InSb, SPIE Proceedings, 836, Conference on Optoelectronic Materials, Devices, Packaging, and Interconnects, 6158-6168.

38. Lischke S; Peczek A; Morgan J; Sun K; Steckler D; Yamamoto Y; Korndorfer F; Mai C; Marschmeyer S; Fraschke M; Kruger A; Beling A; Zimmermann I, Ultra-fast germanium photodiode with 3-dB bandwidth of 265 GHz, Nature Photonics, 2021, 15, 925–931.

39. Li D; Yang Y; Li B; Tang B; Zhang P; Ou X; Sun F; Li Z, High-Speed and High-Power Ge-on-Si Photodetector with Bilateral Mode-Evolution-Based Coupler, Photonics, 2023, 10(2), 142.

40. Moontragoon P; Soref R; Ikonic Z, The direct and indirect bandgaps of unstrained $Si_xGe_{1-x-y}Sn_y$ and their photonic device applications, Journal of Applied Physics, 2012, 112, 073106.

**Acknowledgments** R.S. acknowledges the support of the U.S. Air Force Office of Scientific Research on Grant No. FA9550-21-1-0347. O.M. acknowledges support from NSERC Canada (Discovery, SPG, and CRD Grants), Canada Research Chairs, Canada Foundation for Innovation, Mitacs, PRIMA Québec, Defence Canada (Innovation for Defence Excellence and Security, IDEaS), the European Union's Horizon Europe research and innovation program under grant agreement No 101070700 (MIRAQLS), the U.S. Army Research Office on Grant No. W911NF-22-1-0277, and the U.S. Air Force Office of Scientific and Research on Grant No. FA9550-23-1-0763.

**Declaration of Competing Interest** The authors declare no competing interests.